\documentclass{article}
\usepackage{spconf,amsmath,graphicx}
\usepackage{tabularx,colortbl}
\usepackage{array}
\usepackage[bottom]{footmisc}
\usepackage[ruled,vlined]{algorithm2e}
\usepackage{amsmath}
\usepackage{amsthm}
\usepackage{braket}
\usepackage{multirow}
\usepackage{amsmath}
\usepackage{graphicx}
\usepackage{float}
\usepackage{algorithmicx}
\usepackage{bbm}
 
\newcommand{\norm}[1]{\left\lVert#1\right\rVert}
\usepackage{url}
\usepackage{comment}
\usepackage{movie15}
\usepackage{graphicx}
\usepackage{hyperref}
\usepackage{amssymb}
\usepackage{caption} 
\usepackage{xcolor}
\def\x{{\mathbf x}}

\newtheorem{theorem}{Theorem}

\newif\ifsubmit

\submitfalse
\ifsubmit

\else

\fi
\title{Attributable Watermarking of Speech Generative Models}
\name{Yongbaek Cho\textsuperscript{*}, Changhoon Kim\textsuperscript{*}\thanks{*: Equal Contribution, YC, CK and YY are with the Active Perception Group at the School of Computing and Augmented Intelligence, Arizona State University. YR is with the School for Engineering of Matter, Transport, and Energy, Arizona State University.}, Yezhou Yang, Yi Ren}
\address{Arizona State University, Tempe, AZ, USA}
\begin{document}
%\ninept
%
\maketitle
\begin{abstract}
Generative models are now capable of synthesizing images, speeches, and videos that are hardly distinguishable from authentic contents. Such capabilities cause concerns such as malicious impersonation and IP theft. This paper investigates a solution for model attribution, i.e., the classification of synthetic contents by their source models via watermarks embedded in the contents. Building on past success of model attribution in the image domain, we discuss algorithmic improvements for generating user-end speech models that empirically achieve high attribution accuracy, while maintaining high generation quality. We show the tradeoff between attributability and generation quality under a variety of attacks on generated speech signals attempting to remove the watermarks, and the feasibility of learning robust watermarks against these attacks. Watermarked speech samples are available at \url{https://attdemo.github.io/attdemofull.github.io}.
\end{abstract}
%\href{https://attdemo.github.io/attdemofull.github.io/}{Sound examples}
%

\vspace{-0.1in}
\section{Introduction}
\vspace{-0.1in}
\label{sec:intro}
Generative Adversarial Networks (GANs)~\cite{NIPS2014_5ca3e9b1} have achieved successes in generating artificial contents (e.g., images~\cite{karras2019analyzing}, videos~\cite{clark2019adversarial}, and audios~\cite{gao2018voice}) that are almost indistinguishable from authentic contents. 
These models and their synthetic contents inevitably pose a variety of threats regarding privacy ~\cite{citron2019deepfakes,Satter_2019}, malicious impersonation~\cite{bateman2020deepfakes}, and copyright infringement~\cite{zhang2020not}.
Existing countermeasures to these threats can be categorized into detection~\cite{wang2019cnn} and attribution~\cite{kim2021,yu2020artificial} methods. The detection methods develop binary classifiers to distinguish between generated and authentic contents via intrinsic fingerprints of generative models; the attribution methods develop models from which generated contents are watermarked, so that they can be attributed to their source models via multi-class classification. Recent studies showed that detection may fail when intrinsic fingerprints are removed, e.g., through implicit neural representation~\cite{anokhin2021image}. Instead, 
our focus is attribution, which is much more difficult to spoof. 

\begin{figure}[t]
    \centering
    \includegraphics[width=8cm, height=5cm]{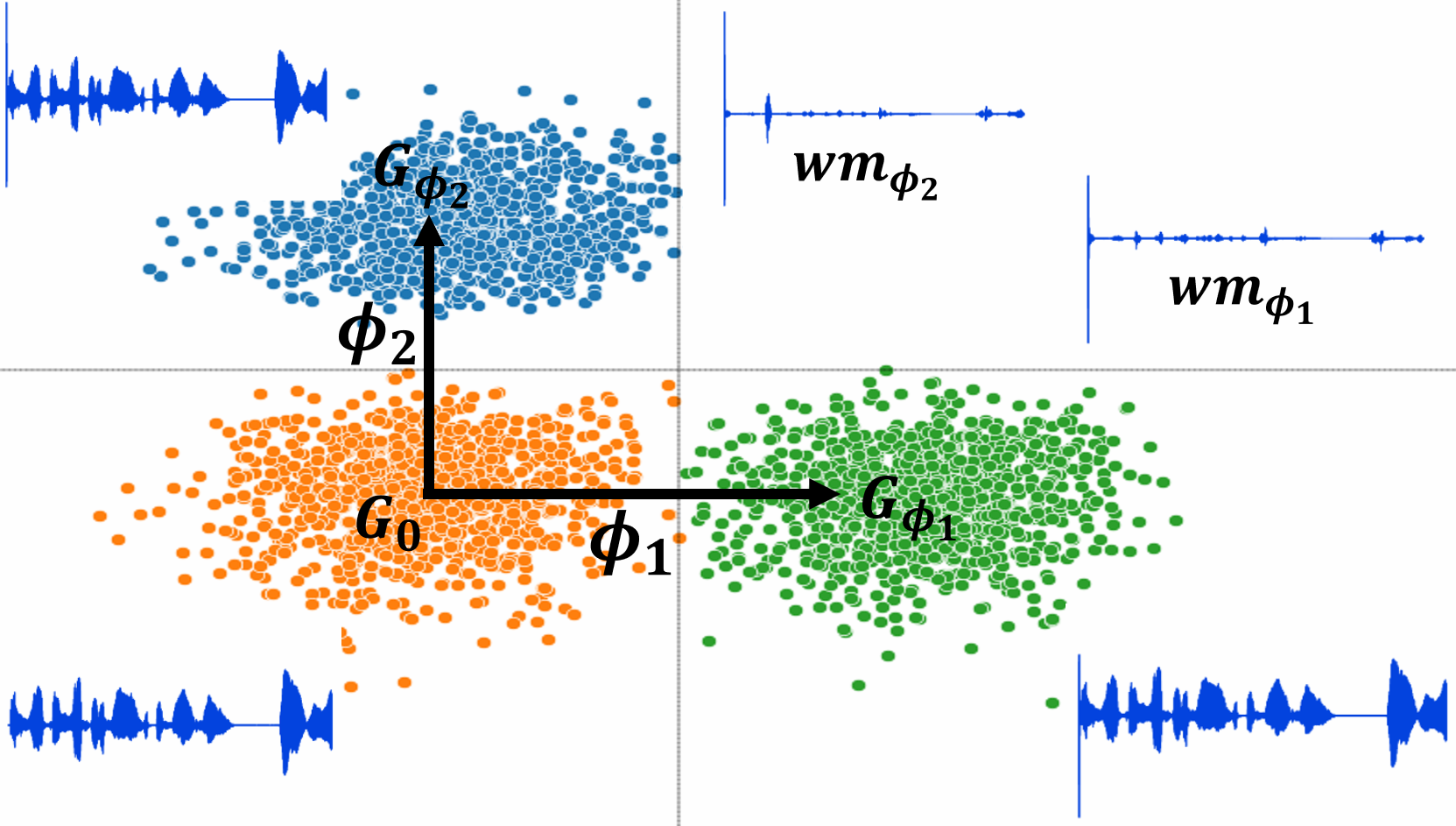}
\vspace{-0.1in}
    \caption{MelGAN model distribution projected to the space spanned by user-specific keys $\phi_1$ and $\phi_2$. The default model $G_0$ is perturbed according to the keys to produce attributable models $G_{\phi_1}$ and $G_{\phi_2}$, which add sparse watermarks ($wm_{\phi_1}$ and $wm_{\phi_2}$) to the beginning of generated speeches.
    \label{fig:concept}}
\vspace{-0.25in}
\end{figure}

\textbf{Protocol:} This study assumes the following model distribution and attribution protocol~(Fig.~\ref{fig:concept}): Consider a model developer who distributes copies of a generative model to its users (e.g., WaveGAN~\cite{donahue2018adversarial} and MelGAN~\cite{kumar2019melgan}). Each user-end model $G_{\phi}: \mathcal{Z} \rightarrow \mathcal{X}$ maps the latent space $\mathcal{Z} \subset \mathbb{R}^{d_z}$ to the content space $\mathcal{X} \subset \mathbb{R}^{d_x}$, and has a key $\phi \in \mathbb{R}^{d_x}$ that defines its unique watermark. A third-party registry (e.g., law enforcement) manages all keys ($\Phi = \{\phi_i\}_{i=1}^N$) and the associated user IDs. The registry accepts contents in question ($x$), performs attribution via a sequence of binary classification, and returns IDs of the users ($i$) who generated the contents~\cite{kim2021} ($\phi_{i}^{T}x > 0$).
% After training speech generative model using our method, the model owner is able to release different copies of the model which include different unique keys but the similar quality of generations. 
% When a user request to download the model, the owner register user's profile with the unique key into the database.
% Then, each user can download the copy of the model with or without knowledge that the key is embedded.
% When misused contents of models are reported, the model owner can collect the misused contents and trace responsible users~\cite{kim2021, yu2020artificial}.
% Following evaluation metrics in the baseline~\cite{kim2021}, 
%Such models pose security and privacy concerns~\cite{citron2019deepfakes,harris2018deepfakes,Satter_2019,breland_2019} as their applications have been found in ***rephrase***
%privacy threats~\cite{***}, malicious impersonation~\cite{DBLP:journals/corr/abs-1802-06840}, and copyright infringement.

\textbf{Sufficient conditions for model attribution}: Within this setting, Kim et al.~\cite{kim2021} studied the sufficient conditions of keys to achieve certifiable attribution. Informally, a set of user-end models are attributable if (1) these models are distinguishable from the authentic dataset, and (2) the inner product of any pair of keys is smaller than a data-dependent threshold. These conditions guide the computation of keys.

%\textbf{Contributions}:
 % We developed new constraints and losses that make our method achieves high evaluation results. We further presented that there is a trade-off between quality of generation and robustness against adversarial post-processes. 
\textbf{Contributions}: We claim the following contributions:
% \begin{enumerate}
% \vspace{-0.05in}
    % \item 
    (1) The algorithm proposed in \cite{kim2021} has only been tested on image generation models. This paper extends the domain to speech generation. We present improvements from \cite{kim2021} to address practical challenges encountered in the speech domain. Specifically, enforcing the alignment between user-end models and their corresponding keys is necessary for maintaining high model attributability. 
% \vspace{-0.05in}
    % \item 
    (2) We empirically test the trade-off between generation quality and robust attributability under adversarial post-processing in the speech domain.

\begin{figure*}[t]
    \centering
    \vspace{-0.1in}
    \includegraphics[width=16cm, height=4cm]{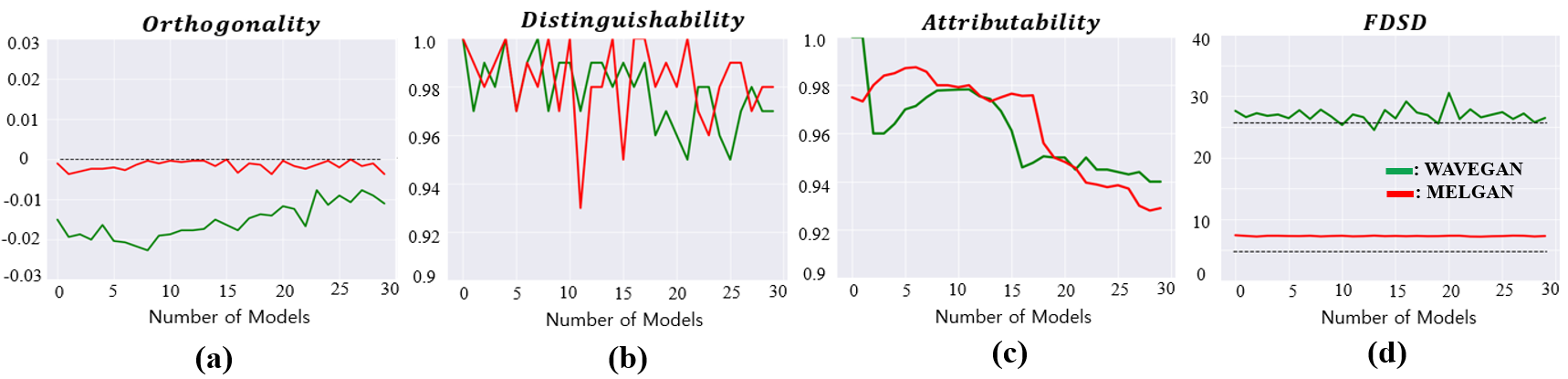}
    \vspace{-0.1in}
    \caption{(a-d) Average orthogonality, distinguishability, attributability, FDSD of 30 WaveGAN user-end models on SC09 and 30 MelGAN user-end models on LJSpeech, respectively. The dotted lines depict baselines.
    \label{fig:metric_graph}}
    \vspace{-0.1in}
\end{figure*}

%%%%%%%%%%%%%%%%%%%%%%%%%%%
% Table 1.                %
%%%%%%%%%%%%%%%%%%%%%%%%%%%
\begin{table}[t]
\caption{Evaluation results for various loss designs.
%We show the results of WaveGAN and MelGAN depending on configurations. 
When proposed configurations are applied, we achieved the best results. 
Aug.:augmented key, Dist.: distinguishability, Att.: attributability, FDSD: Fréchet Deep Speech Distance. $\downarrow$~/~$\uparrow$ indicates lower/higher result is desirable. Base FDSD scores for WaveGAN and MelGAN are 25.65 and 4.74, respectively.\label{tab:configs}}
\centering
\begin{tabular}{l|cccc}
\hline
\multicolumn{1}{l|}{}     & Model   & Dist.$\uparrow$ & Att.$\uparrow$       & FDSD.$\downarrow$ \\ \hline
\multirow{2}{*}{A Baseline~\cite{kim2021}} & WaveGAN & 0.68            & 0.1           & 27.28 \\ \cline{2-5} 
                          & MelGAN  & 0.74          & 0.0              & 12.82     \\ \hline
\multirow{2}{*}{B + Aug.}    & WaveGAN & 0.94          & 0.17          & 30.87     \\ \cline{2-5} 
                          & MelGAN  & 0.99          & 0.68           & 21.85     \\ \hline
\multirow{2}{*}{C + $L_{d}$}     & WaveGAN & 0.97          & 0.31          & \textbf{26.67} \\ \cline{2-5} 
                          & MelGAN  & 0.99          & 0.73           & 7.32    \\ \hline
\multirow{2}{*}{D + $L_{A}$}     & WaveGAN & \textbf{0.98} & \textbf{0.94} & 26.92 \\ \cline{2-5} 
                          & MelGAN  & \textbf{0.99} & \textbf{0.93} & \textbf{7.30}     \\ \hline
\end{tabular}
\vspace{-0.2in}
\end{table}

\vspace{-0.1in}
\section{RELATED WORKS}
\vspace{-0.1in}
\label{sec:related_works}
% \subsection{Audio Generative Models}
\textbf{Speech generative models}:
% Generative models have been successfully demonstrated the fidelity at synthesizing not only images~\cite{} but also in audio~\cite{}. 
GANs~\cite{NIPS2014_5ca3e9b1} were invented to train neural networks which map latent vectors to real distribution via solving real/fake binary classification problem. 
This development brings great success to realistic data synthesis not only in the image domain but also in the speech domain.
Many successful speech synthesis models are established based on GANs (e.g., WaveGAN and MelGAN).
These speech models enable ordinary people to generate realistic fake audio, which warns society against misuse~\cite{Stupp_2019,Metz_2019}.
As one possible solution, our model attribution reveals responsible user who synthesized audio.

%GAN-based speech synthesis models (e.g.,WaveGAN and MelGAN) are indistinguishable by machine or human beings.WaveGAN maps the low-dimensional latent vectors to audio samples.And MelGAN takes a mel-spectrograms generated by text-to-mel transformation as inputs from text-to-mel transformation by a neural network~\cite{shen2018natural}. Such audio generative models are being misused for adversarial purposes~\cite{Stupp_2019}, the risks of which are only amplified through social media~\cite{Metz_2019}.
%GAN-based parallelizable unsupervised audio generative model~\cite{donahue2018adversarial} also manifested the human-recognizable quality.   
%Most generative models~\cite{kumar2019melgan} for text-to-speech synthesis use linguistic features or mel-spectrograms as inputs and back to raw waveforms. 
%***Several connotative and parametric speech synthesis approaches with the application of GANs have been widely deployed from linguistic features to raw waveforms with trained neural vocoder~\cite{}. 
%Autoregressive such as WaveNet \cite{oord2016wavenet} and WaveRNN~\cite{kalchbrenner2018efficient} shows the high perceptual quality and application of GANs for parallelizable unsupervised audio generation~\cite{donahue2018adversarial} manifest the recognizable voice quality.***
% As a response, our methodology discloses responsible user of the misused contents.

% \vspace{-0.1in}
% \subsection{Detection and Attribution of Generative Models}
% \vspace{-0.1in}
\textbf{Detection and attribution of generative models}:
% The advance of generative models led to new societal problems~\cite{DBLP:journals/corr/abs-1802-07228} including malicious personation~\cite{Satter_2019}, fake news~\cite{breland_2019}, and false pornography~\cite{harris2018deepfakes}.
Fake detection~\cite{wang2019cnn} boils down to binary classification between authentic and generated contents. Model attribution, on the other hand, relies on multi-class classification to trace the corresponding models of the generated contents~\cite{marra2019gans,albright2019source,kim2021,yu2020artificial}. Among attribution methods are two directions: model structure attribution~\cite{marra2019gans,albright2019source} and user-end model attribution~\cite{kim2021,yu2020artificial}.
%***The purpose of model structure attribution trains a multi-class classifier that predicts the generator types.***
Model structure attribution is multi-class classification problem of classifying synthesized contents into one of structures of generators (e.g.,~StyleGAN2~\cite{karras2019analyzing}).
However, user-end model attribution classifies contents into responsible user's generator even the users' generators have same structure.
In this work, we focus on user-end model attribution. %(model attribution.
%of fake contents (e.g., StyleGAN~\cite{karras2019style}, StyleGAN2~\cite{karras2019analyzing}, and BigGAN~\cite{DBLP:journals/corr/abs-1809-11096}). 

% On the other hand, user-end model attribution traces the responsible user of a fake contents. 
% This attribution is also defined as multi-class classification problem but the number of classes is the set of users.
% In this work, we are going to focus on user-end model attribution. For brevity, we call it model attribution.
% Since all of the previous research is implemented and studied only in the image domain, we initially propose model attribution methodology in the speech domain.

\vspace{-0.1in}
\section{METHODS}
\vspace{-0.1in}
\label{sec:model_att}
% We present model attribution methodology which enables us to discover responsible user-end model $G_\phi$ of fake speech and to maintain the quality of sound.
% In Sec.~\ref{subsec:problem_state}, we describe how the proposed method can disclose the responsibility of fake content. In Sec~\ref{subsec:key_gen}~and~\ref{subsec:loss_func}, we introduce details of the loss function to achieve model attribution. We follow the training procedure and baseline defined in~\cite{kim2021}. But we will show experimental results that the baseline loss function should be changed~(Tab.~\ref{tab:configs}). 

\subsection{Notations and preliminaries}
\vspace{-0.1in}
\label{subsec:problem_state}
% The indiscriminate advances of generative models are believed to enable the development of malicious applications.
% One promising solution to tackle these problems is to disclose the responsible user of misused contents. 
% We achieve this by distributing user-specific generators which produce realistic speech including user-specific keys.

%Since each speech generative models take a variety of inputs and nueral vocoder based text-to-speech generative models transforms mel-spectrograms to plausible raw wavforms and unsupervised generative models take latent vector as input. Here we define the $s \in$ for unifying all different inputs.
Given an authentic dataset $\mathcal{D} \subset \mathbb{R}^{d_x}$, we assume the existence of a default generator $G_0$ for which the output distribution $P_{G_{0}}$ matches with the authentic data distribution $P_{\mathcal{D}}$. 
% Since the purpose of this paper is not improving the quality, we assume $P_{\mathcal{D}}$ and $P_{G_0}$ are almost same.
Let the user-specific keys be $\Phi := \{\phi_1, \phi_2, ..., \phi_N\}$ for $N$ users, where $\phi_i \in \mathbb{R}^{d_x}$ and $||\phi_i||_2=1$ for $i=1,...,N$.
$G_0$ will be fine-tuned according to all $\phi_i \in \Phi$ to produce the set of user-end generators $\mathcal{G} := \{G_{\phi_{1}},G_{\phi_{2}},...\}$ (see Sec.~\ref{subsec:loss_func}). Let the $i$th binary classifier be $f_{\phi_i}(x) = sign(\phi_{i}^{T}x)$, which returns $1$ if and only if $x \in G_{\phi_i}$. 
% But the difference is the label set of classification can be increasing as generators $\mathcal{G}$ increases.
% We introduce evaluation metrics in the Sec.~\ref{subsec:evaluation}.

%Note that the classifier $f_{\phi_i}(x)$ will return $-1$ even for $x \in P_{G_{0}} \bigcup P_{D}$ because $x \in G_{\phi_i}$.

%We achieve this by training user-end generators $\mathcal{G} := \{G_{\phi_{1}},G_{\phi_{2}},...\}$ with unique keys $\Phi := \{\phi_1, \phi_2, ...\}$ corresponding to each user, such that the owner or the registry of the generative model can trace the responsible user not only with the same generation quality but with distinct keys.
%As a result, the owner can distribute key-embedded model $G_\phi$ instances to the users. When someone abuse generated content, the model owner can trace the responsibility of the user.

We introduce the following metrics to facilitate the discussion: 
(1) \textit{Distinguishability} of $G_{\phi}$ measures the classification accuracy of $f_{\phi}(x)$: 
\vspace{-0.05in}
\begin{equation} \label{eq:distinguishability}
    D(G_{\phi}) := \frac{1}{2}
    \mathbb{E}_{\substack{x \sim P_{G_\phi}\\x_{0} \sim P_{\mathcal{D}}}
    } 
    \left[\mathbbm{1}(f_{\phi}(x)=1)+\mathbbm{1}(f_{\phi}(x_{0})=-1)\right],
    \vspace{-0.05in}
\end{equation}
where $P_{G_{\phi}}$ a user-end distribution. 
% $G$ is $(1-\delta)$-distinguishable for $\delta \in (0,1]$ when $D(G)\geq 1-\delta$. 
(2) \textit{Attributability} measures the averaged classification accuracy of all models of the collection $\mathcal{G} :=\{G_{\phi_1},...,G_{\phi_N}\}$:
\vspace{-0.05in}
\begin{equation} 
    A(\mathcal{G}) := \frac{1}{N}\sum_{i=1}^{N}\mathbb{E}_{x \sim G_{\phi_i}}\mathbbm{1}(\phi_{j}^{T}x < 0, \forall~j\neq i, \phi_{i}^{T}x > 0).
    \label{eq:attributability}
    \vspace{-0.03in}
\end{equation}
% $\mathcal{G}$ is $(1-\delta)$-attributable when $A(\mathcal{G}) \geq 1-\delta$.
(3) 
% We denote by $G_0$
% a default model trained on $\mathcal{D}$, and assume $P_{G_0} = P_{\mathcal{D}}$. 
We measure the \textit{generation quality} of $G_{\phi}$ by Fréchet DeepSpeech Distance~\cite{binkowski2019high}.
%and the $l_1$ norm of the mean output perturbation: 
%\vspace{-0.05in}
%\begin{equation}
%    \Delta x(\phi) = \mathbb{E}_{z\sim P_z}[\norm{G_{\phi}(z;\theta) - G(z;\theta_0)}_1],
%    \label{eq:lack_of_generation_quality}
%\vspace{-0.05in}
%\end{equation}
%where $P_z$ is the latent distribution.

\vspace{-0.05in}
\subsection{Sufficient conditions for model attribution}
\vspace{-0.1in}
\label{subsec:conditions}
We summarize the sufficient conditions for model attribution from \cite{kim2021} in Theorem~\ref{thm:theorem2}.

\begin{theorem}
\label{thm:theorem2} 
We say $\phi$ is data-compliant when $\phi^T x < 0$ for $x \sim P_{\mathcal{D}}$. Let
$d_{\text{min}}(\phi)=\min_{x \in \mathcal{D}} |\phi^T x|$, $d_{\text{max}}(\phi)=\max_{x \in \mathcal{D}} |\phi^T x|$.
% Let $\delta \in [0,1]$. 
% Let 
% \begin{equation}
%     a(\phi,\phi') := -1 + \frac{d_{\text{max}}(\phi') + d_{\text{min}}(\phi') - 2\phi'^T \mu}{\sigma(\phi')\sqrt{\log\left(\frac{1}{\delta^2}\right)} + d_{\text{max}}(\phi') - \phi'^T \mu},
%     \label{eq:rhs}
% \end{equation}
% for keys $\phi$ and $\phi'$.
Then $A(\mathcal{G}) \geq \max\{0, 1- N \delta\}$, if $D(G) \geq 1-\delta$ for all $G_{\phi}\in\mathcal{G}$, and  
\begin{equation}
    \phi^T \phi' \leq \min\left\{\frac{d_{min}(\phi)}{d_{max}(\phi)}, \frac{d_{min}(\phi')}{d_{max}(\phi')}\right\}
    \label{eq:thm2}
\end{equation}
for any pair of data-compliant keys $\phi, ~\phi' \in 
%\mathcal{G}
\Phi
$.
\end{theorem}
From the theorem, model attribution requires keys to be designed in such a way that their corresponding user-end models satisfy (1) data compliance, (2) distinguishability, and (3) the minimal angle constraint in Eq.~\eqref{eq:thm2}. It should be noted that a sufficient and computationally more feasible angle constraint is $\phi^T\phi' \leq 0$.

\vspace{-0.05in}
\subsection{Key generation}
\vspace{-0.1in}
\label{subsec:key_gen}
We now discuss the computation of keys. 
First, we notice that for the tested speech data (SC09~\cite{warden2018speech}, LJSpeech~\cite{ljspeech17}), there does not exist data-compliant keys, i.e., no sub-space classifies the authentic data as one class. This is evident from the low distinguishability in Tab.~\ref{tab:configs}A. We resolve this issue by adding a bias to the binary classifies: $f_{\phi_i}(x) = sign(\phi_{i'}^{T}x+b_i)$ for all $i=1,...,N$. 
The resultant distinguishability improves as seen in Tab.~\ref{tab:configs}B.
To reduce notational burden, we will denote $[\phi'_i, b_i]$ by $\phi_i$ and the augmented data (with appended 1s) by $x$.
% To match dimension of $x$ and $\phi$, we added constant into additional dimension of $x$.
%when calculating $f_{\phi}(x)$.

Each new key is generated by solving the following problem with existing keys $\phi_j$ for $j=1,...,i-1$:
\vspace{-0.05in}
\begin{equation}
\label{eq:key_generation}
    \min_{\phi} \mathbb{E}_{x \sim G_{0}} 
    [\max\{1+f_{\phi}(x), 0\}] 
    + \sum_{j=1}^{i-1}\max\{\phi_{j}^T \phi, 0 \}.
    % \phi_{i} = \arg\min_{\phi} \mathbb{E}_{x \sim G_{0}} 
    % [\max({1+f_{\phi}(x), 0})] 
    % + \sum_{j=1}^{i-1}\{\phi_{j}^T \phi, 0 \},
    %\phi^{T}\begin{bmatrix} x \\ 1\end{bmatrix}
\vspace{-0.05in}
\end{equation}
The orthogonality penalty (second term of RHS) is omitted for the generation of the first key ($i=1$).
% By solving this equation, we can train keys that are mutually orthogonal and satisfies data compliance.
% We note that the set of keys $\Phi$ is not fixed. If needs more keys, the owner can train additional key which satisfies eq.~\eqref{eq:key_generation}.
\begin{table*}[t]
\centering
\small
\caption{
Evaluation metrics before (Bfr.) and after (Afr.) robust training against adversarial post-processes. Before robust training, FDSD scores of WaveGAN and MelGAN are 26.92 and 7.30, respectively~(Tab.~\ref{tab:configs}D).
Dist. = Distinguishability, Att. = Attributability \label{tab:robust_result}}
\vspace{-0.1in}
\begin{tabular}{cccccccccccc}
\hline
Metric                 & Model   & \multicolumn{2}{c}{Noise} & \multicolumn{2}{c}{Gain} & \multicolumn{2}{c}{Speed} & \multicolumn{2}{c}{Pass filter} & \multicolumn{2}{c}{Combination} \\
                       &         & Bfr.        & Afr.        & Bfr.        & Afr.       & Bfr.        & Afr.        & Bfr.           & Afr.           & Bfr.           & Afr.           \\ \hline
\multirow{2}{*}{Dist. $\uparrow$} & WaveGAN & 0.91        & 0.98        & 0.95        & 0.98       & 0.85        & 0.98        & 0.94           & 0.98           & 0.79           & 0.92           \\
                       & MelGAN  & 0.97        & 0.99        & 0.88        & 0.97       & 0.60           & 0.86        & 0.80           & 0.99           & 0.73           & 0.95           \\ \hline
\multirow{2}{*}{Att. $\uparrow$}  & WaveGAN & 0.88        & 0.96        & 0.94        & 0.98       & 0.71        & 0.90        & 0.64           & 0.91           & 0.31           & 0.73           \\
                       & MelGAN  & 0.72        & 0.92        & 0.63        & 0.88       & 0.40        & 0.70        & 0.64           & 0.84           &  0.23              & 0.56                \\ \hline
\multirow{2}{*}{FDSD. $\downarrow$} & WaveGAN & \multicolumn{2}{c}{36.54} & \multicolumn{2}{c}{42.58}     & \multicolumn{2}{c}{46.12}  & \multicolumn{2}{c}{50.85}            & \multicolumn{2}{c}{47.56}            \\
                       & MelGAN  & \multicolumn{2}{c}{7.99}      & \multicolumn{2}{c}{8.55}     & \multicolumn{2}{c}{24.48}      & \multicolumn{2}{c}{9.415}            & \multicolumn{2}{c}{27.49}            \\ \hline
\end{tabular}
\vspace{-0.1in}
\end{table*}

\vspace{-0.05in}
\subsection{Retraining of user-end generator}
\vspace{-0.1in}
\label{subsec:loss_func}
While Theorem 1 holds when $G_{\phi}$ models the perturbed distribution $\{x + \phi | x\in P_{\mathcal{D}}\}$, this exact match of distributions may not be achieved in practice due to the limited capacity of $G_{\phi}$ and the domain-specific boundaries of $x$ (e.g., for speech data, $x \in [-1,1]^{d_x}$). We found through experiments that this mismatch deteriorates the attributability of user-end models. In the following, we describe a practical formulation for retraining the default model $G_0$ so that the resultant user-end model $G_{\phi}$ will (1) be distinguishable from the authentic data, (2) have low generation quality drop, and (3) be attributable. 
% Tab.~\ref{tab:configs} shows the necessity of proposed configurations. \\

\textbf{Distinguishability loss}: We use a standard hinge loss to penalize $G_{\phi}$ on distinguishability:
\vspace{-0.05in}
\begin{equation}
     L_{h} = \mathbb{E}_{x \in P_{G_{\phi}}}
     \max\{1 - f_{\phi}(x),0\}.
\vspace{-0.05in}
\end{equation}

\textbf{Generation quality loss}: 
%Our goal is to minimize the quality degradation and we compute the Mean absolute error loss between the key-embedded user-end generative models and the original fake models.
To discourage quality degradation, we introduce a loss that computes the expected distance between samples from the user-end and the default models. 
We utilize $l_1$ distance which gives better perceptual quality than $l_2$ distance~\cite{pandey2018adversarial}:
\vspace{-0.05in}
\begin{equation}
    L_{d} = \mathbb{E}_{z \sim P_{z}} \left[\norm{G_{0}(z) - G_{\phi}(z)}_{1}\right].
\vspace{-0.05in}
\end{equation}

\textbf{Angle loss}:
%Training user-end generative models with conventional losses included hinge loss and Mean Abolute loss(MAE) showed the high attributability in the image domain~\cite{}. 
% Training user-end generative models using $L_{h}$ and $L_{d}$ guarantee the high attributability in the image domain~\cite{kim2021} but this is not the case in speech domain.
Through experiments, we notice that the expected perturbation $\mathbb{E}_{z \sim P_{z}}\left[G_{\phi}(z) - G_0(z)\right]$ may not align with $\phi$, which causes attirbutability to be lower than expected. See Tab.~\ref{tab:configs}C.
Therefore, we propose an angle loss to encourage the alignment:
\begin{equation}
    L_{A} = \max\left\{1 - { (G_{0}(z)-G_{\phi}(z)) \cdot \phi \over \norm{(G_{0}(z)-G_{\phi}(z))}_2 \cdot \norm{\phi}_2}, 0\right\}.
\end{equation}
%Tab. x reports the metrics of distinguishability and attributability after model attribution training with using angular loss or not.
Tab.~\ref{tab:configs}D shows the effectiveness of the angle loss at improving the empirical attributability of 30 user-end models.
The training objective is thus the following:
\vspace{-0.05in}
\begin{equation} \label{func:userend}
     \min_{G_{\phi}}~ 
    \lambda_{1} L_{h} + \lambda_{2} L_{d} + \lambda_{3} L_{A},
\vspace{-0.05in}
\end{equation} 
where $\lambda_{1}$,$\lambda_{2}$ and $\lambda_{3}$ are set to 10, 10000, 1000,  respectively. We optimize this loss function to create $G_{\phi_i}$ iteratively.

\begin{figure*}[t]
    \centering
    \includegraphics[width=17cm, height= 5.5cm]{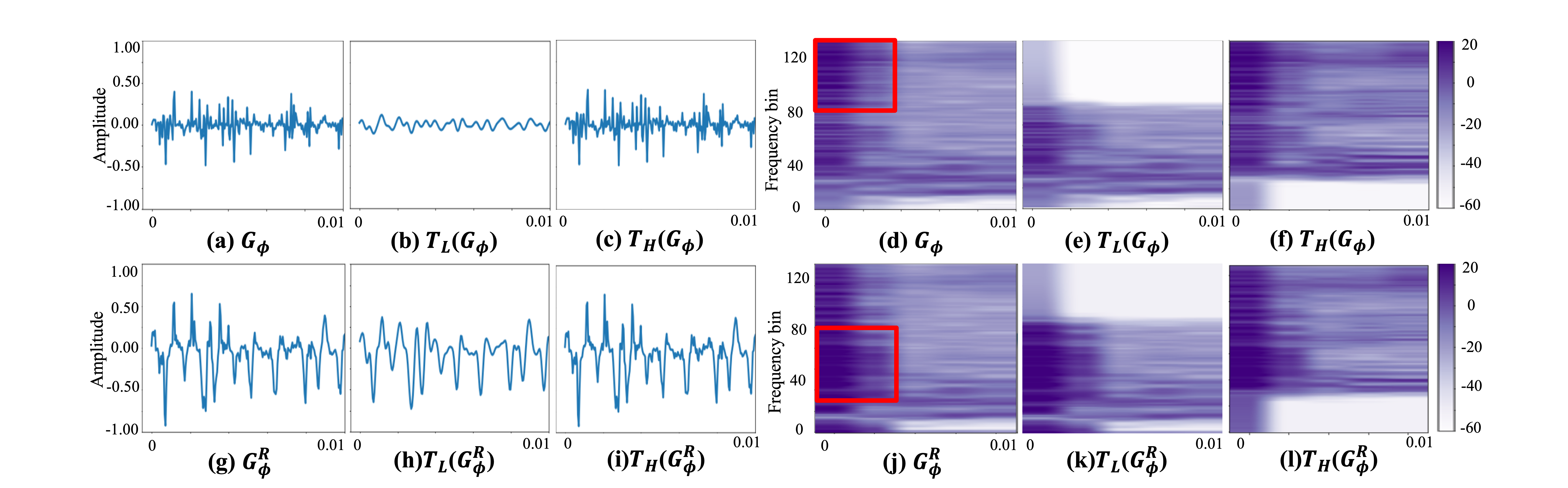}
    \vspace{-0.1in}
    \caption{Results of robust training against pass-filter attack.
    (a,g): Audio signal from a non-robust generator $G_\phi$ and the corresponding robust generator $G_{\phi}^{R}$. (d,j): Corresponding Mel-spectrogram of (a,g). The peak-amplitude regions are highlighted. (b,c,h,i,e,f,k,l): Audio signal and Mel-spectrogram after the attack.
    $T_{L}$~/~$T_{H}$ indicates low and high pass-filter, respectively.
    \label{fig:wav_robustness}
    }
\vspace{-0.2in}    
\end{figure*}

\vspace{-0.1in}
\section{EXPERIMENTS}
\vspace{-0.1in}
\label{sec:majhead}
% In this section, we introduce experimental details and results.
% We describe the setup, evaluation metrics, and results in Sec.~\ref{subsection:experimental_set}, Sec.~\ref{subsec:evaluation}, and Sec.~\ref{subsec:exp_results}, respectively. Finally, we concern about adversarial post-processing in Sec.~\ref{subsection:adversarial}.\\

% \vspace{-0.1in}
\subsection{Experimental setup}
\vspace{-0.1in}
\label{subsection:experimental_set}
\textbf{Dataset}:
We tested our model attribution using SC09~\cite{warden2018speech} and LJspeech~\cite{ljspeech17} datasets.
SC09 is a subset of speech commands by a variety of speakers that include spoken ten vocabulary words from zero to ten each of a duration of 1 second. The dataset is split into training, test and validation sets consisting of 18.5k, 2.5k, and 2.5k data points, respectively. LJspeech contains 13.1k audio clips by a single speaker. We split LJspeech into 11.5k, 0.5k, 0.5k for training, test, and validation, respectively.
\textbf{Model and training}:
%In our experiments, all models were trained on NVIDIA TESLA V100 GPUs for the LJSpeech at 22,050 Hz and SC09 at 16,384 Hz.
WaveGAN maps the latent vectors to audio samples and we directly employed SC09 dataset to train. MelGAN takes a mel-spectrograms and we cut each LJspeech clip to 3 seconds in length.

\vspace{-0.1in}
\subsection{Experimental results}
\vspace{-0.1in}
\label{subsec:exp_results}
We report in Fig.~\ref{fig:metric_graph} the average distinguishability, attributability, and the average generation quality (in terms of FDSD) of a sequence of user-end WaveGAN and MelGAN models being iteratively trained by solving Eq~\eqref{eq:key_generation} and Eq.~\eqref{func:userend}. Results with all 30 models are reported in Tab.~\ref{tab:configs}D.
% All models are trained following steps proposed in the Sec.~\ref{sec:model_att}.
% We experimentally demonstrate that improving the baseline configuration by adding a bias, angular loss, and $l_{1}$ distance metric instead of $l_2$~(Tab.~\ref{tab:configs}).
% We found that the distinguishability was remarkably improved in WaveGAN and MelGAN almost 26\% and 25\%, respectively~(Tab.~\ref{tab:configs}(B)).
% Augmented dimension with using vectors as a bias input enables to keys to be achieved the data compliant. 
%As can be observed from~Fig.~\ref{fig:metric_graph}, the orthogonality of keys learned by augmented feature vector as a bias input helps to be linearly separable and distinguishable by decision boundary given by keys.
%Augmented dimension enables keys to linearly separate authentic distribution $P_{G_{0}}$ and user-end models' distribution $P_{G_\phi}$. 
It should be highlighted that the angle loss significantly improves the attributability of models, achieving 94\% and 93\% on WaveGAN and MelGAN, respectively. 
% trained user-end models with baseline configuration is not attributable, since trained user-end models $G_{\phi}$ are not aligned with the specific key, 
This shows that in practice, it is necessary to align the trained model $G_{\phi}$ with the corresponding $\phi$. 
% In a nutshell, our angular loss adjusts the angle between $G_{\phi}$ and specific key $\phi$ so that make the high attributability be attainable.
%Supervised by Angular loss, $x \in G_{\phi_i}$ is angulary distributed and able to be attributable. 
% Regarding generation quality, trained $G_{\phi}$ with $L_{d}$ shows notable quality improvements comparing to baseline configurations quantitatively~(Tab.~\ref{tab:configs}(C)).

%We trained 30 user-specific models to  

\vspace{-0.1in}
\subsection{Adversarial post-processing}
\vspace{-0.1in}
\label{subsection:adversarial}

Lastly, we test the robustness of our method against various post-processes that aim at removing the watermarks from generated contents.
Following the experimental setting in~\cite{yu2018attributing,yu2020artificial,kim2021}, we assume that the registry is aware of the distribution of attacks $P_T$, where $T: \mathbb{R}^{d_x} \rightarrow \mathbb{R}^{d_x}$ can represent (1) adding noise, (2) gain, (3) changing speed, (4) combined pass filters, and the combination of these four. 
% We randomly apply all post-processing with 50\%. 
To train robust user-end models $G_{\phi}^{R}$, we propose the following problem formulation for updating user-end models given $\phi$:
\vspace{-0.05in}
\begin{equation}
\begin{aligned}
     \min_{G_{\phi}} 
     & ~\mathbb{E}_{{T \sim P_T, ~x \sim P_{G_{\phi}}}} 
     \left[\lambda_1 \max\big\{1-f_{\phi_{i}}(T(x),0)\big\}\right. \\
     & ~ \left. + \lambda_{2} L_{d} + \lambda_{3} L_{A}\right].
     %+ C[\norm{G_{0}(m) - %G_{\phi_{i}}(m;\theta_{i})}_{1} \right]}
\vspace{-0.05in}
\end{aligned}
\end{equation}
%To assess the trade off between robustness of user-end model and its quality, we report the results after robust training by post-processing and evaluations are given in Tab.~\ref{tab:robust_result}. 

\textbf{Setup}: Details of post-processes are as follows. \textit{Noise}: Noise type is uniformly sampled from Brown, Blue, Violet, and Pink Noise.
% Here output of user-end generative models multiplied by a random amplitude factor and this would increase or reduce the volume. 
\textit{Gain}: Gain multiplies a random amplitude to reduce or increase the volume.
%\textit{Gain}: We uniformly sample gain from $[-18, 6]$dB. 
\textit{Pass filter}: High and low pass filters are both considered. We set the cut off frequency to $[2200, 4000]$ for low pass filter and $[200, 1200]$ for high pass filter, respectively. It should be noted that the frequency ranges are chosen to avoid removing the semantic contents of the generated speeches. \textit{Speed}: Speed perturbations speed up or slow down an audio signal with re-sampling.
% Warped time signal will affect the frames of utterances. 
The speed percentage is uniformly chosen from $[80, 90, 110]$. Lastly, \textit{combination} attacks combine the other four attacks, each with a 50\% chance to be applied.
\textbf{Results}: Tab.~\ref{tab:robust_result} reports the average distinguishability, attributability, and generation quality with and without robust training against post-processes.
% Detailed settings of post-processing are provided in Sec.~\ref{subsection:adversarial}. 
A tradeoff is observed between robust attributability and generation quality.
%In our experiments, we consider five types of post-processes: noise, gain, pass-filter, speed change, and combination of these four. 
%Result of Robust training
%Robustness and generation quality are summarized in Tab.~\ref{tab:robust_result}. 
%We show the results of the generative models which are post-processed by adversaries. We considered five attack types which are widely used (e.g., Noise, Gain, Speed and Combined Pass filter). 
% \vspace{-0.1in}
% \section{Discussion}
% \vspace{-0.1in}
% \textbf{Robust Watermark:} 
% For real world applications, we hope our methodology is robust to various post-processes. While quantitative results of robust training are available in Tab.~\ref{tab:robust_result}, visual watermark was not given. 
To further understand the effectiveness of robust training, here we pick low/high-pass filters as the attack and compare a non-robust watermark and its corresponding robust version for MelGAN in Fig.~\ref{fig:wav_robustness}a,g, as well as their filtered watermarks in  Fig.~\ref{fig:wav_robustness}b,c,h,i. We focus on the first 0.01 second of the signals where watermarks dominate. From the results, we see that robust training successfully finds watermarks that have frequency ranges in between the low- and high-pass filters.
To further support this finding, we average Mel-spectrogram before and after filters over 1000 samples in Fig.~\ref{fig:wav_robustness}(d-f, j-l). We reiterate that since attacks should avoid removing the semantic contents of a generated speech, there always exists a frequency window for which robust watermarks can be created.

\vspace{-0.1in}
\subsection{Collusion Attack}
\vspace{-0.1in}
%On one side, There are several oblivious threats against collusion attacks in real world applications. 
We define a collusion attack as to merge multiple sources of the same content to produce a new copy that averages source watermarks and potentially makes it difficult to attribute ~\cite{su2000capacity}. 
However, this attack will not be successful with the presented definition of attributability. To explain, attacker(s) download two models of the collection $G_{1}$ and $G_{2}$, and interpolate the output by $x_{new} = \lambda x_{1} + (1-\lambda) x_{2}$, where $x_{i} \in G_{\phi_i} \forall i \in \{1,2\}$, and $\lambda \in [0,1]$. 
When $G_{1}$ and $G_{2}$ are attributable, for any key $\phi$ that does not belong to $G_{1}$ or $G_{2}$,  we have $\phi^Tx_1 <0$ and $\phi^Tx_2 <0$. Thus, $\phi^{T} (\lambda x_1 + (1-\lambda) x_2) < 0$.

%However, this attack will not be successful with the presented definition of attributability. To explain, attacker(s) download two models of the collection $\mathcal{G} :=\{G_{\phi_1},...,G_{\phi_N}\}$, and interpolate the output by $x_{new} = \lambda_1 x_{1} +...+ \lambda_n x_{n}$, where $x_{i} \in G_{\phi_i}$, and $\sum_i \lambda_i = 1~\forall \lambda_i \geq 0$. While maintaining the requirements Eq.(2), for any key $\phi$ that does not belong to attributed $\mathcal{G} :=\{G_{\phi_1},...,G_{\phi_N}\}$. Thus, $\phi^{T}(\lambda x_{1} +...+ (1-\lambda)x_{n}) < 0$.

\vspace{-0.1in}
\section{Conclusion}
\vspace{-0.1in}
\label{sec:Conclusion}
We investigated the feasibility of model attribution in the speech domain. Our method is based on a protocol where the model distributor trains attributable user-end generative models by embedding unique watermarks.
We showed that in practice, it is necessary to enforce the alignment between user-end models and their designated keys in order to achieve empirically high attributability in the speech domain. This is verified on WaveGAN and MelGAN trained on SC09 and LJSpeech datasets, respectively. Lastly, we revealed the tradeoff between generation quality and robust attributability.

\vspace{-0.1in}
\section{ACKNOWLEDGMENTS}
\vspace{-0.1in}
This work is partially supported by the National Science Foundation under Grant No. 2101052 and by an Amazon AWS Machine Learning Research Award (MLRA). Any opinions, findings, and conclusions expressed in this material are those of the author(s) and do not reflect the views of the funding entities.

\newpage
\bibliographystyle{IEEEbib}
\bibliography{IEEEbib}
\end{document}